# Coexistence of Quasi-two-dimensional Superconductivity and Tunable Kondo Lattice in a van der Waals Superconductor

Shiwei Shen[1], Tian Qin[1], Jingjing Gao[2,3], Chenhaoping Wen[1], Jinghui Wang[1,4], Wei Wang[2,3], Jun Li[1,4], Xuan Luo[2], Wenjian Lu[2], Yuping Sun[2,5,6,*], Shichao Yan[1,4,*]

[1] *School of Physical Science and Technology, ShanghaiTech University, Shanghai 201210, China*

[2] *Key Laboratory of Materials Physics, Institute of Solid State Physics, HFIPS, Chinese Academy of Sciences, Hefei 230031, China*

[3] *University of Science and Technology of China, Hefei 230026, China*

[4] *ShanghaiTech Laboratory for Topological Physics, ShanghaiTech University, Shanghai 201210, China*

[5] *High Magnetic Field Laboratory, HFIPS, Chinese Academy of Sciences, Hefei 230031, China*

[6] *Collaborative Innovation Centre of Advanced Microstructures, Nanjing University, Nanjing 210093, China*

*\*Email: yanshch@shanghaitech.edu.cn; ypsun@issp.ac.cn*

## Abstract

Realization of Kondo lattice in superconducting van der Waals materials not only provides a unique opportunity for tuning the Kondo lattice behavior by electrical gating or intercalation, but also is helpful for further understanding the heavy fermion superconductivity. Here we report a low-temperature and vector-magnetic-field scanning tunneling microscopy and spectroscopy study on a superconducting compound (4Hb-TaS$_2$) with alternate stacking of 1T-TaS$_2$ and 1H-TaS$_2$ layers. We observe the quasi-two-dimensional superconductivity in the 1H-TaS$_2$ layer with anisotropic response to the in-plane and out-of-plane magnetic fields. In the 1T-TaS$_2$ layer, we detect the Kondo resonance peak that results from the Kondo screening of the unpaired electrons in the Star-of-David clusters. We also find the intensity of the Kondo resonance peak is sensitive to its relative position with the Fermi level, and it can be significantly enhanced when it's further shifted towards the Fermi level by evaporating Pb atoms onto the 1T-TaS$_2$ surface. Our results are not only important for fully understanding the electronic properties of 4Hb-TaS$_2$, but also pave the way for creating tunable Kondo lattice in the superconducting van der Waals materials.

## Main text

Superconductivity in two-dimensional materials has spurred great research interests in the field of solid-state physics, including the Ising superconductivity in monolayer transition metal dichalcogenides (TMDs) and the unconventional superconductivity in twisted-bilayer graphene.[1-3] The coupling between the two-dimensional superconductors and layered magnetic materials leads to the discovery of many interaction-driven electronic phases, such as topological superconductivity in monolayer $CrBr_3$ grown on the superconducting TMDs material 2H-$NbSe_2$.[4] Recently, localized spins have been observed in the charge density wave (CDW) superlattice of 1T-phase single-layer group-V TMDs materials, such as single-layer 1T-$TaS_2$, 1T-$TaSe_2$ and 1T-$NbSe_2$, and the Kondo lattice emerges when they couple to the two-dimensional metallic layers.[5-7] Stacking the group-V TMDs layers hosting Kondo lattice with the two-dimensional superconductors would provide a new platform for exploring the interplay between superconductivity and Kondo lattice.

4Hb-$TaS_2$ is a superconducting van der Waals material consisting of an alternate stacking of 1T-$TaS_2$ and 1H-$TaS_2$ layers [Fig. 1(a)],[8,9] which in their bulk form host the correlated and superconducting electronic phases, respectively.[10-12] During cooling down, 4Hb-$TaS_2$ goes through two charge density wave (CDW) transitions. The 1T layers in 4Hb-$TaS_2$ enter the $\sqrt{13} \times \sqrt{13}$ CDW state with periodic Star-of-David (SD) clusters during the CDW transition at ~315 K, and the $3 \times 3$ CDW superlattice appears in the 1H layers below 22 K.[13,14] 4Hb-$TaS_2$ becomes superconducting when the temperature is below ~2.7 K [Fig. 1(b)]. In comparison with the transition temperature of bulk 2H-$TaS_2$ ($T_c$ = 0.7 K), the enhancement of $T$c is likely due to the charge transfer between the 1T and 1H layers in 4Hb-$TaS_2$, which performs electron doping to the 1H layers.[15-17] During the superconducting phase transition, the signs of time-reversal symmetry breaking have been observed in the previous muon relaxation rate measurements, which suggests the possible chiral superconductivity in 4Hb-$TaS_2$.[8] Although the recent low-temperature STM experiment demonstrates the evidence of topological boundary modes in the superconducting state of 4Hb-$TaS_2$, the origin of the time-reversal-symmetry breaking is still unclear.[8,9]

In this work, we use low-temperature and vector-magnetic-field scanning tunneling microscopy/spectroscopy (STM/STS) to study the superconducting and magnetic properties of 4Hb-$TaS_2$. Our measurements reveal the quasi-two-dimensional superconductivity in the 1H layer and the Kondo resonance in the 1T layer coexist in the 4Hb-$TaS_2$. We also demonstrate that the strength of the Kondo resonance peak can be tuned by changing its relative position with the Fermi level.

*Experimental.* Single crystals of 4Hb-$TaS_2$ were synthesized with chemical vapor transport method using iodine as the transport agent, and the detailed grown conditions were described elsewhere.[13] 4Hb-$TaS_2$ samples were cleaved in an ultrahigh-vacuum chamber and then immediately inserted into the STM head for measurements. For dosing Pb, Pb atoms were deposited onto the 4Hb-$TaS_2$ sample from a Knudsen cell, and the 4Hb-$TaS_2$ sample was kept at room temperature. After dosing Pb, the 4Hb-$TaS_2$ sample was kept at room temperature for ~30 minutes and then inserted into the STM head. The STM measurements were conducted by using a home-built He-3 STM with vector magnet. The tungsten tips were flashed by electron-beam bombardment for several minutes before use. The d$I$/d$V$ spectra were measured using a standard lock-in detection technique.

*Results.* Cleaving the 4Hb-$TaS_2$ sample results in large-area flat surfaces with 1T or 1H termination (Supplemental Section 1). The two terminations can be distinguished by their CDW patterns shown in the constant-current STM topography. As shown in Figs. 1(c) and 1(d), the 1T layer shows strong $\sqrt{13} \times \sqrt{13}$ CDW pattern with periodic SD clusters, and the 1H layer hosts the CDW superlattice with $3 \times 3$ periodicity (More bias-voltage-dependent STM topographies taken on the 1T and 1H layers are shown in Supplemental Sections 2 and 3).[9,18] In the differential conductance (d$I$/d$V$) spectrum taken at 0.7 K, we detect the superconducting gap in the 1H layer, but no clear superconducting gap is observed in the 1T layer [Fig. 1(e)].[9] Although there is a dip-like feature near the Fermi level in the d$I$/d$V$ spectrum taken on the 1T layer [Fig. 1(e)], this dip has no response to the magnetic field, which indicates it is not related with the superconductivity (Supplemental Section 4). The absence of superconducting gap in the 1T layer could be due to the mismatch between the Fermi surfaces in the 1H and 1T layers, the proximity induced superconducting gap in the 1T layer is much smaller than the gap in the 1H layer.[8] This also indicates the residual heat capacity in 4Hb-TaS$_2$ below $T$c may be due to the non-superconducting 1T layers.[8]

The superconducting gap in the 1H layer can be well fitted with the Dynes model with a finite density of states offset, yielding a gap of ~0.32 meV [Fig. 1(e)], which agrees with the recently reported STM results.[9,19] The finite density of states offset in the fitting may come from the contribution from the underneath non-superconducting 1T-TaS$_2$ layer. The superconducting gap in the 1H layer can be suppressed by increasing temperature to be above the $T$c [Fig. 1(e)]. Interestingly, we find that the superconducting gap in the 1H layer has different responses to the in-plane and out-of-plane magnetic fields. As shown in Figs. 1(f) and 1(g), the superconducting gap can be fully suppressed with 0.5 T out-of-plane magnetic field, and it can still be detected even with 1.5 T in-plane magnetic field indicating the high in-plane critical field. Our STS results demonstrate that 4Hb-TaS$_2$ is a quasi-two-dimensional superconductor and the superconductivity is from the 1H layers, which is consistent with the recent transport and low-temperature STM measurements on 4Hb-TaS$_2$.[8,9]

Having characterized the quasi-two-dimensional superconductivity in the 1H layer of 4Hb-TaS$_2$, we then focus on the low-temperature electronic properties of the 1T layer. In our previous STM data taken at 4.3 K (above $T$c), we have shown that in the 1T layer there is a narrow electronic band near the Fermi level that originates from the unpaired electrons in the SD clusters.[16,20] Due to the charge transfer between the adjacent 1T and 1H layers, the narrow electronic band in the 1T layer is shifted to be slightly above the Fermi level [Fig. 2(a)].[16] In the d$I$/d$V$ spectra taken at low temperature (0.7 K) and low bias modulation (1 mV or less), the fine structures in the narrow band are observed [Fig. 2(c)], and these features can be missed when performing d$I$/d$V$ measurements with large modulation voltage. Fig. 2(c) shows the high-resolution d$I$/d$V$ spectra taken along the purple arrow in Fig. 2(b). As we can see, there is a small V-shaped gap with the minimum at the Fermi level, and there are also three electronic peaks located above the Fermi level which are marked by the red, orange, and yellow arrows. In the d$I$/d$V$ spectra taken in different regions of the 1T layers (Supplemental Section 5), the energy positions of these electronic peaks have spatial variations [Fig. 2(d)], which could be due to the slightly different electronic coupling between the 1T layer and the underneath 1H layer. As shown in Fig. 2(d), when these peaks are located closer to the Fermi level, the relative intensity of the first electronic peak that is closest to the Fermi level becomes higher.

In order to further tune the energy positions of these electronic peaks, we evaporate Pb atoms onto the 4Hb-TaS$_2$ sample, which could perform electron doping effect to the 1T layer.[21] Pb atoms form regular islands on both the 1T and 1H layers. While there are individual randomly distributed Pb atoms on the bare 1H layer, almost no Pb atoms can be seen on the bare 1T layer [Fig. 3(a) and Supplemental Section 6]. In comparison with the d$I$/d$V$ spectrum taken on the pristine 1T layer, after dosing Pb atoms the narrow band near the Fermi level is slightly shifted towards the Fermi level and a sharp resonance peak emerges at the Fermi level [Fig. 3(c)]. We also find that the shift of the narrow electronic band is not directly induced by the Pb islands, but is because of the intercalated Pb atoms below the 1T layer which appear as dark SD clusters in the constant-current STM topography (see Supplemental Section 7 for more details).[21] Fig. 3(d) is the line cut d$I$/d$V$ spectra taken along the purple arrow in Fig. 3(b). We can see that when the electronic peaks marked by the arrows are shifted further towards the Fermi level, the intensity of the first electronic peak is significantly enhanced.

Recently, in the 1T layer of the epitaxially-grown 1T-TaS$_2$ and 1H-TaS$_2$ heterostructure, the Kondo resonance peak at the Fermi level has been observed,[5] and the first electronic peak located closest to the Fermi level in Figs. 2(d) and 3(d) is likely the Kondo resonance peak resulting from the Kondo screening of the unpaired electrons in the SD clusters. To confirm the Kondo nature of the enhanced electronic peak near the Fermi level, we perform temperature and magnetic-field dependent d$I$/d$V$ measurements. As shown in Fig. 3(e), the Kondo resonance peak has strong temperature dependence. At each temperature, we fit the Kondo resonance peak with a thermally convolved Fano lineshape and the electronic peaks located above the Kondo resonance with the Gaussian functions (see Supplemental Section 8 for more details). As shown in Fig. 3(f), the half width at half maximum of the temperature-dependent Kondo peak can be fitted with the well-known Kondo expression $\frac{1}{2}\sqrt{(\alpha k_B T)^2 + (2 k_B T_K)^2}$,[6,22,23] which yields a Kondo temperature $T_K \approx 26$ K. The Kondo resonance peak is significantly broadened at higher temperature (above 25 K) and its intensity is reduced, which makes the Kondo resonance peak strongly overlap with the nearby electronic peaks [Fig. 3(e)]. In the magnetic-field-dependent d$I$/d$V$ spectra, the Kondo resonance peak is slightly broadened as increasing the magnetic field, but no clear peak splitting is detected with magnetic field up to 8.5 T [Fig. 3(g) and Supplemental Section 9]. We note that the Kondo resonance peak with narrower width can be clearly split with ~ 10 T magnetic field in the 1T layer of the epitaxially grown 1T/1H TaS$_2$ heterostructure.[5] We think larger magnetic field is needed to further split the Kondo resonance peak shown in Fig. 3(g) (Supplemental Section 10).

Figure 4(a) summarizes the d$I$/d$V$ spectra shown in Fig. 2(d) and the typical d$I$/d$V$ spectra in Fig. 3(d), which clearly demonstrate the Fermi-level tuning of the Kondo resonance peak. In order to quantify the enhancement of the Kondo resonance in the 1T layer of 4Hb-TaS$_2$, we first fit the Kondo resonance peak and the nearby electronic peaks with thermally convolved Fano lineshape and Gaussian functions, respectively [Fig. 4(a) and Supplemental Section 8]. Then we plot the ratio between the intensities of the Kondo peak and the nearest electronic peak as a function of the energy position of the Kondo peak [Fig. 4(b)]. As we can see from Fig. 4(b), the relative intensity of the Kondo resonance peak increases exponentially as it moves towards the Fermi level. This effect is also consistent with the Anderson model where the Kondo temperature $T_K \sim \exp(-1/J)$ with the Kondo coupling constant $J$. When the Kondo peak shifts towards the Fermi level, the Kondo coupling constant and the Kondo temperature increase.[24]

*Discussion and conclusion.* Our STS data indicate that the tunable electronic peak located near the Fermi level is consistent with the Kondo resonance peak resulting from the Kondo screening of the unpaired electrons in the SD clusters, which forms the Kondo lattice in analog to that in the heavy fermion system. In our case, the narrow electronic band derived from the unpaired electrons in the SD clusters of the 1T layer replaces the *f*-electron band in the heavy fermion system, and the itinerant band comes from the underneath 1H layer. We also find that when the Kondo resonance peak is shifted towards the Fermi level, its intensity can be strongly enhanced. This kind of Fermi-level tuning of the Kondo resonance has also been observed in the alloy system Y$_{1-x}$U$_x$Pd$_3$, where the U f-electron spectral weight is shifted toward Fermi level and Kondo resonance emerges.[24,25] This further confirms the Kondo nature of the tunable resonance peak. As shown in Fig. 4(a), in the 1T layer of 4Hb-TaS$_2$, the dip-like feature with the minimum at the Fermi level gradually disappears as the Kondo resonance peak shifts towards the Fermi level. This indicates that this dip-like feature may be due to the interference of the tunneling channels to the Kondo peak and the conduction band.[26] Furthermore, in comparison with the d$I$/d$V$ spectrum taken on the heavy fermion metals,[27,28] the electronic peaks marked by the orange and yellow arrows in Figs. 2(c) and 3(d) are likely due to the crystal-electric-field excitations in the 1T layer of 4Hb-TaS$_2$.[27-29]

In the heavy fermion systems, the coupling between the Kondo resonance mode and the itinerant electrons leads to the hybridized electronic gap.[28,30] Recently, the gap-like feature is observed in the 1H layer of the epitaxially grown 1T/1H TaS$_2$ heterostructure,[5] which has been attributed to be the heavy fermion hybridization gap. We note that there is no superconductivity and no $3 \times 3$ CDW superlattice in the 1H layer of the 1T/1H TaS$_2$ heterostructure.[5] In the d$I$/d$V$ spectrum taken on the 1H layer of 4Hb-TaS$_2$, except the superconducting gap, there is also a gap-like feature near the Fermi level and it persists when the superconducting gap is fully suppressed by the magnetic field (Supplemental Section 11). Since the Kondo resonance in the pristine 1T layer of 4Hb-TaS$_2$ is weak, this gap is likely not the hybridization gap, and it is the CDW gap in the 1H layer.

Finally, we discuss the possible connection between the Kondo resonance in the 1T layer and the enhanced muon relaxation rate at the superconducting transition temperature of 4Hb-TaS$_2$.[8] It could be that the local moments in the 1T layer also contribute to the muon relaxation. During the superconducting transition, the superconducting gap opens in the 1H layer and the density of states near the Fermi level is reduced. This could weaken the Kondo screening of the local moments in the 1T layer and affect the muon spin relaxation rate.[8]

In summary, we demonstrate the quasi-two-dimensional superconductivity and the Kondo resonance peak coexist in 4Hb-TaS$_2$. The strength of the Kondo resonance peak is sensitive with its relative position with the Fermi level, and the Kondo resonance can be greatly enhanced when it is tuned toward the Fermi level. The discovery of the Kondo resonance in the 1T layer is important for understanding the transport properties of 4Hb-TaS$_2$.[8,13,31] We also believe that the 1T-TaS$_2$-related superconducting materials may provide a new platform for exploring the interplay between Kondo resonance and superconductivity, which would be helpful for further understanding the heavy fermion superconductivity.[30,32-35]

*Acknowledgments*. The authors thank Prof. Patrick A. Lee for fruitful discussions. S.Y. acknowledges the financial support from the National Key R&D Program of China (Grant No. 2020YFA0309602), National Science Foundation of China (Grant No. 11874042) and the start-up funding from ShanghaiTech University. C.W. acknowledges the support from National Natural Science Foundation of China (Grant No. 12004250) and the Shanghai Sailing Program (Grant No. 20YF1430700). J.W. acknowledges the support from National Natural Science Foundation of China (Grant No. 12004251) and the Shanghai Sailing Program (Grant No. 21YF1429200). J.G., W.W., X.L., W.L and Y.S. thank the support of National Key Research and Development Program under Contract No. 2021YFA1600201, the National Nature Science Foundation of China under Contracts No. 11674326 and No. 11774351, and the Joint Funds of the National Natural Science Foundation of China and the Chinese Academy of Sciences' Large-Scale Scientific Facility under Contracts No. U1832141, No. U1932217 and U2032215.

Figure 1

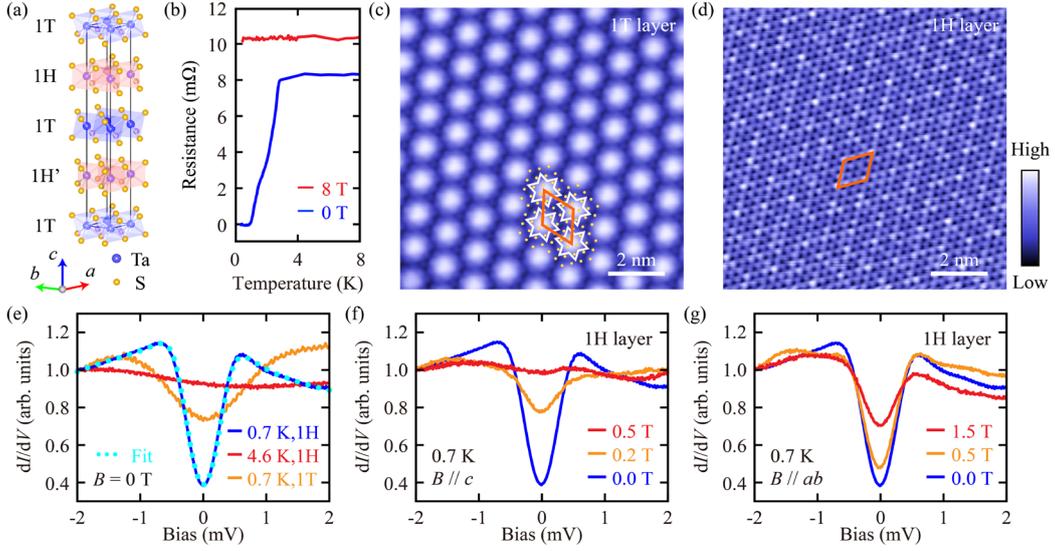

**Fig. 1.** Quasi-two-dimensional superconductivity in the 1H layer of 4Hb-TaS$_2$. (a) Schematic for the structure of 4Hb-TaS$_2$ with alternate stacking of the 1T and 1H layers. (b) Temperature-dependent resistance of 4Hb-TaS$_2$ with 0 T and with 8 T out-of-plane magnetic fields. The superconducting transition temperature $T_c$ defined by the onset point is ~2.7 K. (c) and (d) Constant-current STM topographies taken on the 1T layer (c: $V_s$ = 300 mV, $I$ = 300 pA) and 1H layer (d: $V_s$ = 10 mV, $I$ = 20 pA). The orange rhombuses in (c) and (d) indicate the $\sqrt{13} \times \sqrt{13}$ and the $3 \times 3$ CDW unit cells, respectively. (e) The d$I$/d$V$ spectra taken on the 1H layer at 0.7 K (blue) and 4.6 K (red), and on the 1T layer at 0.7 K (yellow). The dotted line is the superconducting gap fitted with the Dynes model and an offset. (f) Out-of-plane magnetic field dependent d$I$/d$V$ spectra taken on the 1H layer. (g) In-plane magnetic field dependent d$I$/d$V$ spectra taken on the 1H layer.

**Figure 2**

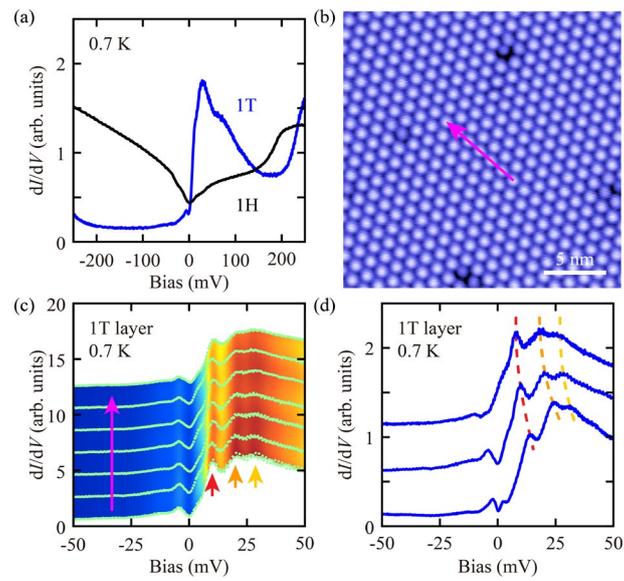

**Fig. 2.** Electronic structure in the 1T layer of 4Hb-TaS$_2$. (a) The d$I$/d$V$ spectra taken on the 1H (black) and 1T (blue) layers at 0.7 K. (b) Constant-current STM topography taken on the 1T layer ($V$s = 100 mV, $I$ = 50 pA). (c) Line cut of d$I$/d$V$ spectra taken along the purple arrow in (b). The red, orange, and yellow arrows indicate three fine electronic peaks located above the Fermi level. (d) The typical d$I$/d$V$ spectra taken on different regions of the 1T layer. The spectra are vertically offset for clarity.

# Figure 3

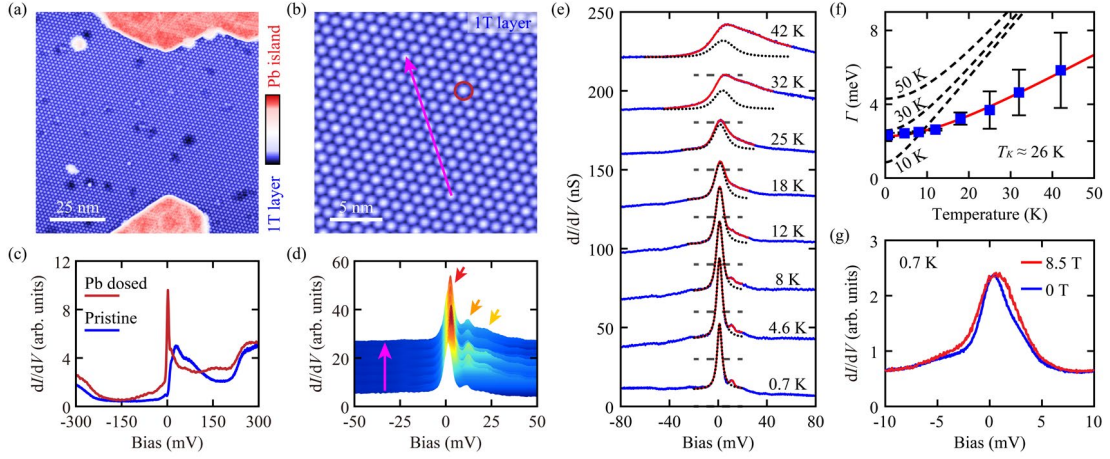

**Fig. 3.** Tuning the Kondo resonance in the 1T layer. (a) Constant-current STM topography with Pb island on the 1T layer of 4Hb-TaS$_2$ ($V_s = -200$ mV, $I = 20$ pA). (b) Constant-current STM topography taken on the bare 1T layer region ($V_s = -200$ mV, $I = 50$ pA). (c) The d$I$/d$V$ spectrum (red) taken on the circles in (b) and the d$I$/d$V$ spectrum (blue) taken on the pristine 1T layer before dosing Pb. (d) Line cut of the d$I$/d$V$ spectra taken along the purple arrow in (b) at 0.7 K. The red, orange, and yellow arrows in (d) indicate three electronic peaks. (e) The blue lines are the d$I$/d$V$ spectra measured in the 1T layer at different temperatures. The red lines are the thermally convolved Fano fits to the Kondo resonance peak (the black dotted spectra) and the Gaussian fits to the nearby electronic peaks. The spectra are vertically offset for clarity. (f) Evolution of the measured half width at half maximum ($\Gamma$) of the Kondo resonance peak with temperature (blue squares with error bar). The black dashed lines indicate the function $\frac{1}{2}\sqrt{(\alpha k_B T)^2 + (2k_B T_K)^2}$ with $\alpha = 2\pi$ and $T_K = 10, 30$ and $50$ K. The red line is the fit with $\alpha = 3.1$, which yields a Kondo temperature of $T_K \approx 26$ K. (g) The d$I$/d$V$ spectra measured with 0 T (blue) and with 8.5 T (red) out-of-plane magnetic fields.
9

**Figure 4**

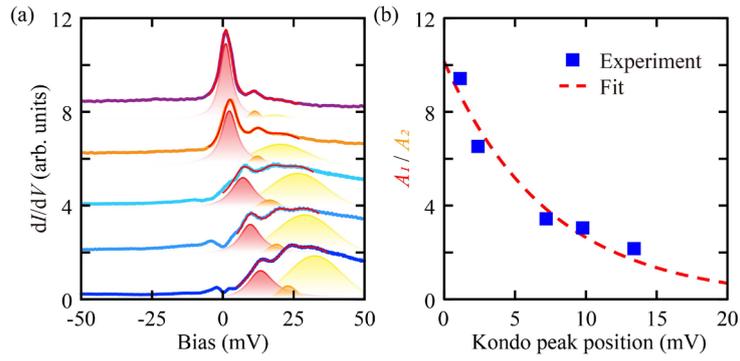

**Fig. 4.** Fermi-level tuning of the Kondo resonance in the 1T layer of 4Hb-TaS$_2$. (a) Summary of the d$I$/d$V$ spectra shown in Figs. 2(d) and 3(d). The red lines are the fits of the d$I$/d$V$ spectra where the Kondo resonance peak is fitted with the thermally convolved Fano lineshape and the other two electronic peaks are fitted with the Gaussian functions. The spectra are vertically offset for clarity. (b) The blue squares are the relative intensities of the Kondo resonance peak (the ratio between the intensities of the Kondo resonance peak and the nearest electronic peak) in the d$I$/d$V$ spectra shown in (a) as a function of the position of the Kondo peak. The red dashed line is the exponential fit to the experimental data.

# SUPPLEMENTARY MATERIALS

# Coexistence of Quasi-two-dimensional Superconductivity and Tunable Kondo Lattice in a van der Waals Superconductor

Shiwei Shen[1], Tian Qin[1], Jingjing Gao[2,3], Chenhaoping Wen[1], Jinghui Wang[1,4], Wei Wang[2,3], Jun Li[1,4], Xuan Luo[2], Wenjian Lu[2], Yuping Sun[2,5,6,*], Shichao Yan[1,4,*]

[1] *School of Physical Science and Technology, ShanghaiTech University, Shanghai 201210, China*

[2] *Key Laboratory of Materials Physics, Institute of Solid State Physics, HFIPS, Chinese Academy of Sciences, Hefei 230031, China*

[3] *University of Science and Technology of China, Hefei 230026, China*

[4] *ShanghaiTech Laboratory for Topological Physics, ShanghaiTech University, Shanghai 201210, China*

[5] *High Magnetic Field Laboratory, HFIPS, Chinese Academy of Sciences, Hefei 230031, China*

[6] *Collaborative Innovation Centre of Advanced Microstructures, Nanjing University, Nanjing 210093, China*

*Email: yanshch@shanghaitech.edu.cn; ypsun@issp.ac.cn

1. **STM topographies taken on the 1T and 1H layers of 4Hb-TaS$_2$**

   As shown in Fig. S1, cleaving 4Hb-TaS$_2$ sample results in large-area surfaces with 1T or 1H surface termination.

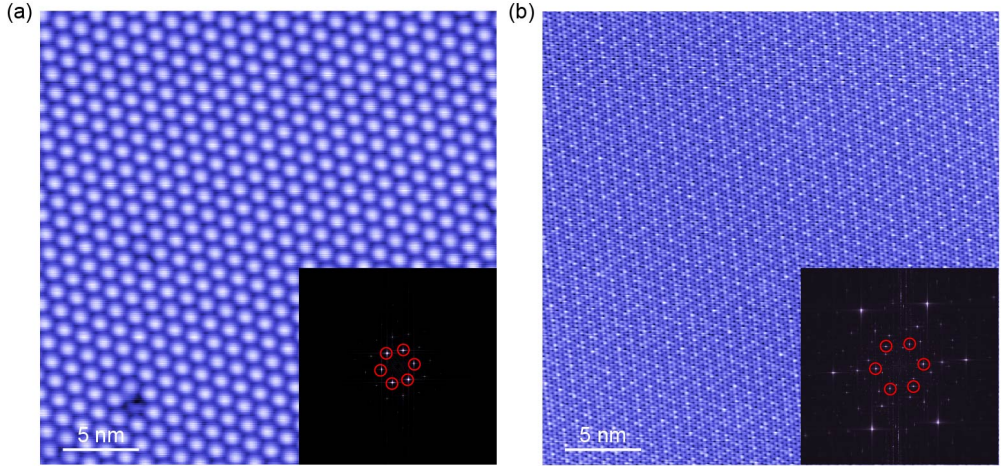

**Figure S1. a,** Constant-current STM topography taken on the 1T layer of 4Hb-TaS$_2$ ($V$s = 500 mV, $I$ = 20 pA). **b,** Constant-current STM topography taken on the 1H layer of 4Hb-TaS$_2$ ($V$s = 10 mV, $I$ = 20 pA). The insets are the Fourier transform images of (a) and (b), respectively. The red circles in the insets indicate the $\sqrt{13} \times \sqrt{13}$ CDW wavevectors in (a) and $3 \times 3$ CDW wavevectors in (b).

2. **Bias-voltage dependent STM topographies taken on the 1T layer of 4Hb-TaS$_2$**

   Constant-current STM topographies taken on the 1T layer show the periodic SD clusters.

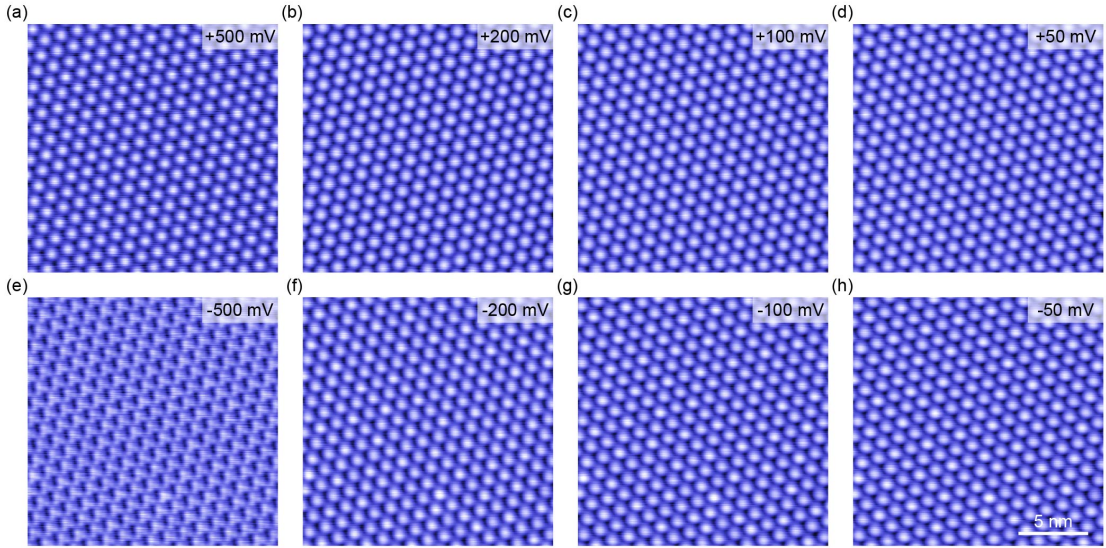

**Figure S2. a-h,** Constant-current STM topographies taken in the same region of the 1T layer with various bias voltages.

### 3. Bias-voltage dependent STM topographies taken on the 1H layer of 4Hb-TaS$_2$

As shown in bias-voltage-dependent STM topographies taken on the 1H layer, the 3 × 3 CDW superlattice can be clearly seen in the STM topographies taken with low-bias-voltage STM topographies (Figs. S3d and S3h). In the STM topographies taken with high-bias voltages, the imprint of the $\sqrt{13} \times \sqrt{13}$ CDW superstructure from the underneath 1T layer appears (Figs. S3a and S3e).

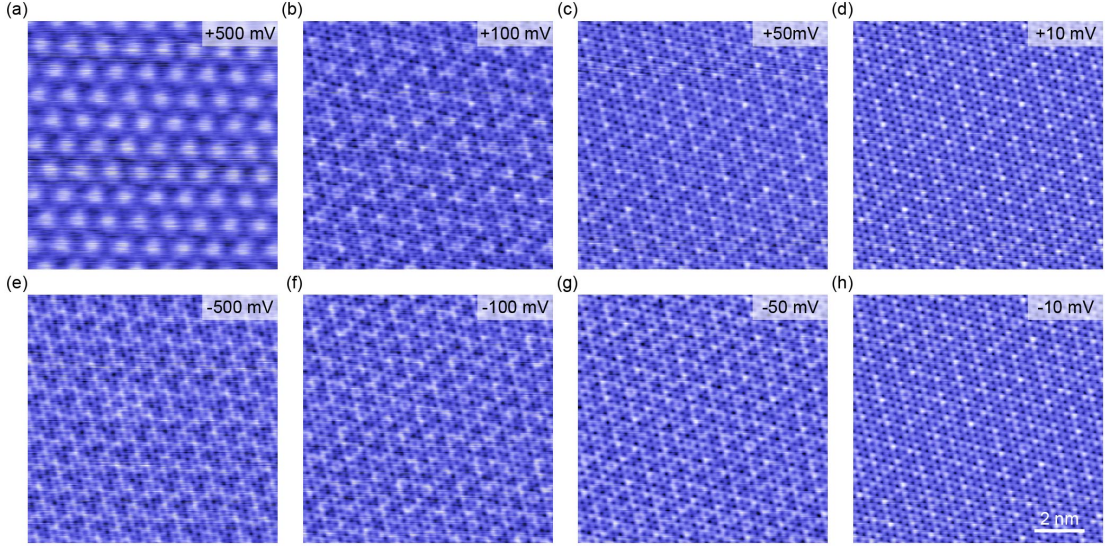

**Figure S3. a-h,** Constant-current STM topographies taken in the same region of the 1H layer with various bias voltages.

### 4. d$I$/d$V$ spectra taken on the 1T layer of 4Hb-TaS$_2$ without and with magnetic field

As we can see from the d$I$/d$V$ spectra taken on the 1T layer, there is a dip near the Fermi level which has no response to the external magnetic field. This indicates that this dip is not the superconducting gap.

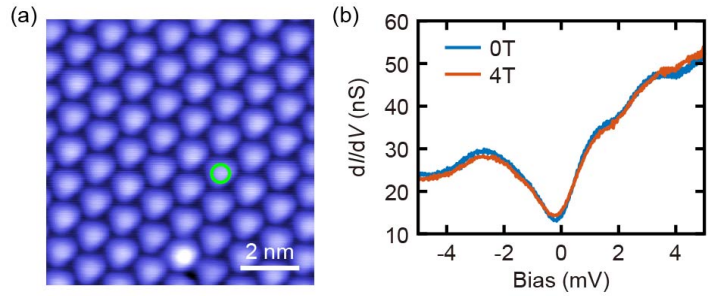

**Figure S4. a,** Constant-current STM topography taken on the 1T layer of 4Hb-TaS$_2$ ($V_s = -500$ mV, $I = 20$ pA). **b,** d$I$/d$V$ spectra taken on the position shown by the green circle in (a) with 0 T (blue) and 4 T (orange) out-of-plane magnetic fields at 0.7 K.

## 5. d*I*/d*V* spectra taken in the different regions of the 1T layer

Figure S5 shows the STM topographies taken on the 1T layer where the d*I*/d*V* spectra in Fig. 2d are taken. In different regions of the 1T layer, the relative positions of the three electronic peaks and the Fermi level are slightly different, which could be due to the slightly different interlayer coupling with the underneath 1H layer.

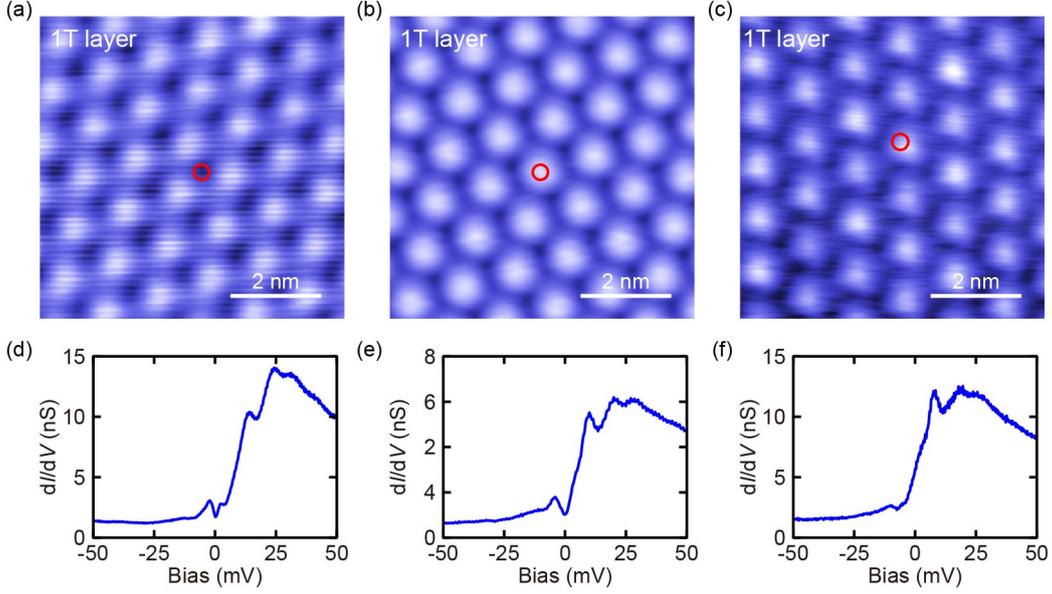

**Figure S5. a-c,** Constant-current STM topography taken in the different regions of the 1T layer of 4Hb-TaS$_2$ (a: $V_s = -200$ mV, $I = 20$ pA; b: $V_s = 100$ mV, $I = 50$ pA; c: $V_s = -500$ mV, $I = 20$ pA). **d-f,** d*I*/d*V* spectra taken at the positions shown by the red circles in (a)-(c).

## 6. Dosing Pb atoms onto the 1T and 1H layers of 4Hb-TaS$_2$

As shown in Fig. S6, Pb islands are formed on both the 1T and 1H layers of 4Hb-TaS$_2$. However, individual randomly distributed Pb atoms can be found in the 1H layer, and no Pb atoms can be seen in the bare 1T surface. As we can see from the line profile across the step edge (inset of Fig. S6a), the height difference between the bare 1T layer and the individual Pb atoms on the 1H layer is ~500 pm. As we can see in Figs. S6b and S6d, the strong Kondo resonance appears in the bare 1T layer.

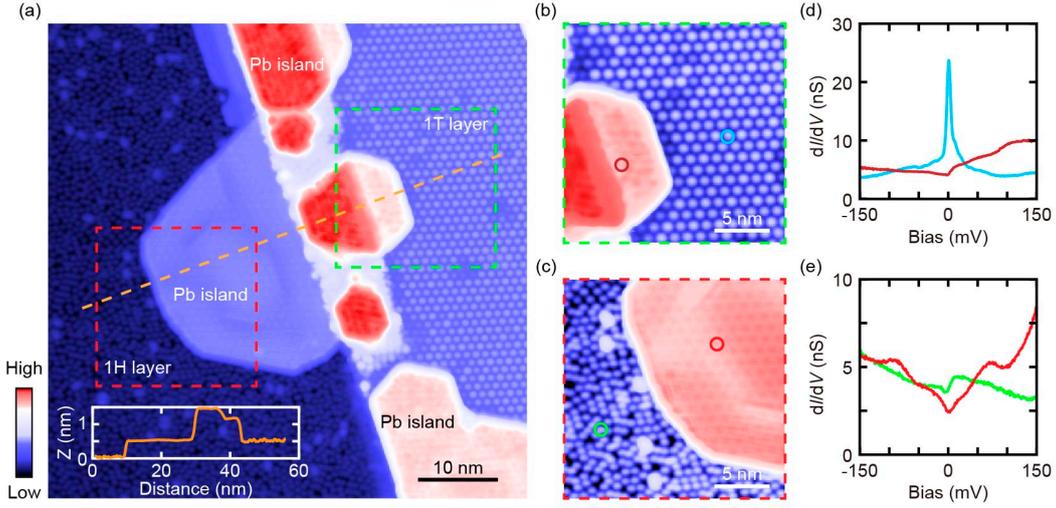

**Figure S6. a,** Constant-current STM topography taken near the boundary between the 1H layer and 1T layer with Pb islands ($V_s = -200$ mV, $I = 20$ pA). The inset is the line profile along the orange dashed line. **b and c,** Constant-current STM topographies taken on green and red dashed squares, respectively ($V_s = -200$ mV, $I = 20$ pA). **d and e,** $dI/dV$ spectra taken on colored circles in (b) and (c), respectively.

## 7. Doping effect to the 1T layer after evaporating Pb atoms onto 4Hb-TaS$_2$

Figure S7 shows the $dI/dV$ spectra taken in the regions near and far from the Pb islands. As we can see that the $dI/dV$ spectra taken in the region near the Pb island (the purple spectra in Fig. S7b) are similar as the $dI/dV$ spectra taken in the region far from the Pb island (the red spectra in Fig. S7c). This indicates that doping effect after evaporating Pb atoms onto the 1T layer is not directly from the Pd islands. Otherwise, we would expect the doping effect is stronger in the region that is closer to the Pb islands.

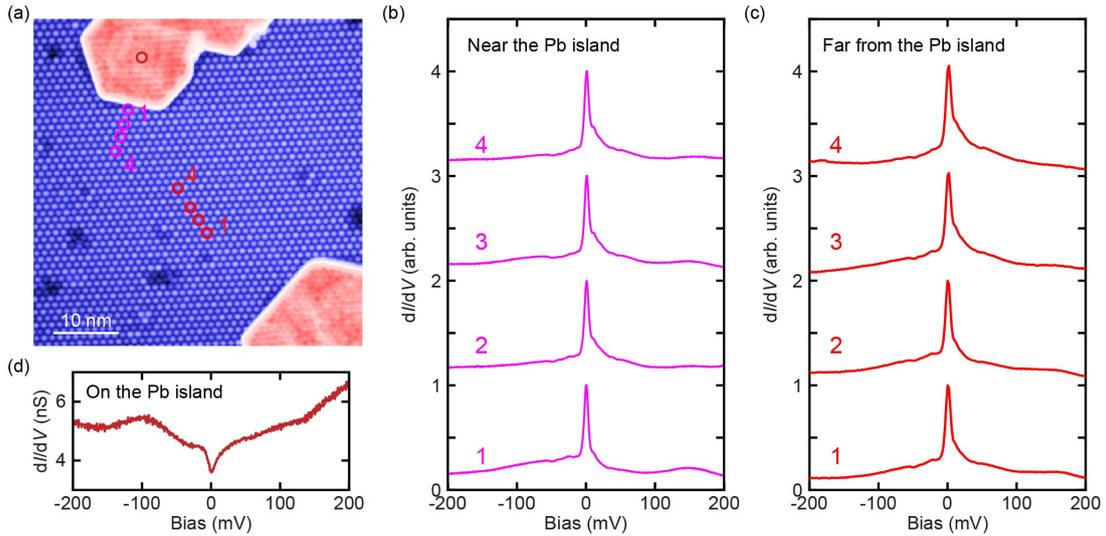

**Figure S7. a,** Constant-current STM topography on the 1T layer with Pb islands ($V_s = -200$ mV, $I = 20$ pA). **b,** $dI/dV$ spectra taken on the points marked by the purple circles in (a). **c,** The $dI/dV$ spectra taken on the points marked by the red circles in (a). **d,** $dI/dV$ spctrum taken on the Pb island marked by the dark red circle in (a).

As shown in Figs. 4a and S8a, after evaporating Pb atoms, there are several dark SDs in the bare 1T-TaS$_2$ region. In the d$I$/d$V$ spectra taken on the dark SD clusters (Fig. S8b), the narrow electronic band is shifted to be slightly above the Fermi level, which indicates the SD clusters are induced by the intercalated Pb atoms. Similar effect has also been observed in Ref. 21.

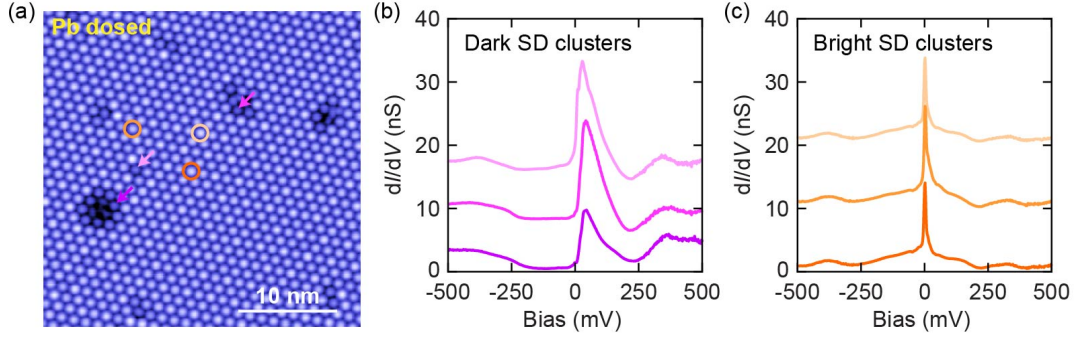

**Figure S8**. **a,** Constant-current STM topography in the bare 1T-TaS$_2$ region after dosing Pb atoms ($V$s = −200 mV, $I$ = 50 pA). **b,** d$I$/d$V$ spectra taken on the points marked by the colored arrows in (a). **c,** d$I$/d$V$ spectra taken on the points marked by the colored circles in (a).

## 8. Fitting the Kondo resonance peak and the nearby electronic peaks

To extract the half width at half maximum (HWHM) of the Kondo peak, the Kondo peak in the d$I$/d$V$ spectrum is fitted with a thermally convolved Fano lineshape.

$$\frac{dI}{dV}(eV) = \int_{-\infty}^{\infty} \left( A_1 \cdot \frac{(\xi + q)^2}{1 + \xi^2} + C \right)(-f'(E - eV, T)) dE, \quad \xi = \frac{E - E_0}{\Gamma}$$

Here, the first term in the integrand is a scaled Fano function plus a constant background, and the $f(E, T) = 1/(e^{E/(k_B T)} + 1)$ is the Fermi distribution function. In the Fano function, $q$ is the Fano parameter, $\Gamma$ is the HWHM of the Kondo peak, and $E_0$ is the energy position of the Kondo peak.

To fit the d$I$/d$V$ spectra shown in Figs. 3e and 4a, we fit the Kondo peak and the two electronic peaks located at higher energy. The Kondo peak is fitted with a thermally convolved Fano lineshape and the two nearby electronic peaks are fitted with two Gaussian functions.

$$\frac{dI}{dV}(eV) = \int_{-\infty}^{\infty} \left( A_1 \cdot \frac{(\xi + q)^2}{1 + \xi^2} + C + A_2 \cdot e^{-\frac{(E - \mu_2)^2}{2\sigma_2^2}} + A_3 \cdot e^{-\frac{(E - \mu_3)^2}{2\sigma_3^2}} \right)(-f'(E - eV, T)) dE$$

Here the first term in the integrand includes a scaled Fano function with a constant background and two Gaussian functions. In the Gaussian functions, $A_2$ and $A_3$ are the amplitudes, $\mu_2$ and $\mu_3$ are the energy positions, and $\sigma_2$ and $\sigma_3$ are the standard deviations of the functions.

## 9. Magnetic field dependence of the Kondo resonance peak

As shown in Fig. S9, the width of the Kondo resonance peak gradually increases as increasing the magnetic field. However, no clear splitting is observed with magnetic field up to 8.5 T.

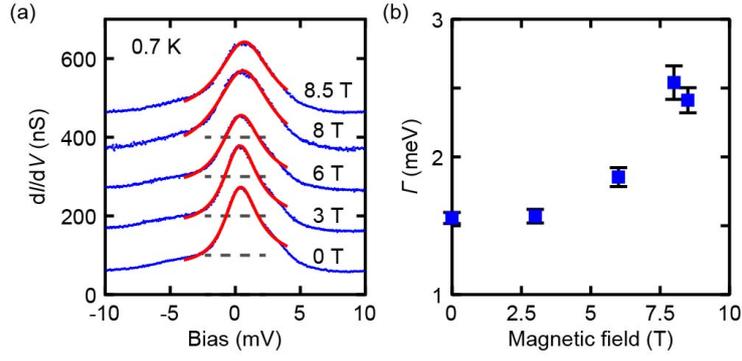

**Figure S9. a,** The blue lines are the d$I$/d$V$ spectra measured with different out-of-plane magnetic fields at 0.7 K. The red lines are the thermally convolved Fano fits to the electronic peak at the measured magnetic fields (see Section 8 for more details). The spectra are vertically offset for clarity. **b,** Half width at half maximum ($\Gamma$) of the Kondo resonance peaks as a function of the magnetic field.

## 10. Comparison between the Kondo resonance peak in the 1T/1H-TaS$_2$ heterostructure (Ref. 5) and the Kondo resonance peak in Fig. 3g.

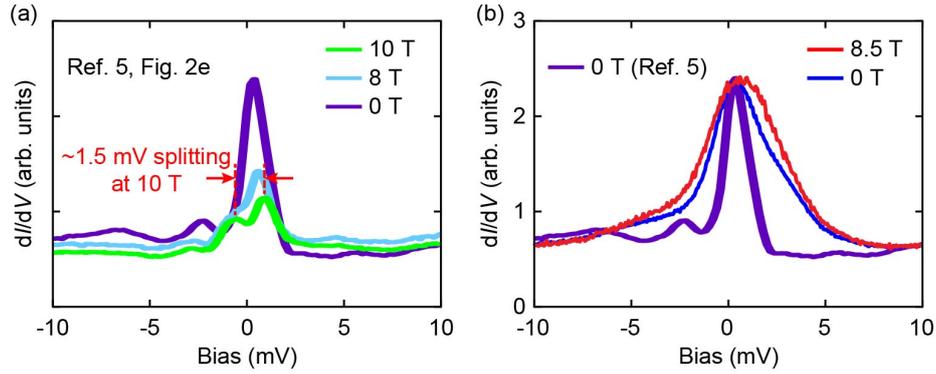

**Figure S10. a,** The d$I$/d$V$ spectra taken on the 1T layer of the 1T/1H-TaS$_2$ heterostructure with 0 T (purple), 8 T (blue) and 10 T (green) magnetic fields (from Ref. 5). (b) The d$I$/d$V$ spectra taken with 0 T (blue) and 8.5 T (red) magnetic fields shown in Fig. 3g. The purple spectrum is the same spectrum as the purple spectrum in (a).

## 11. Gap-like feature in the 1H layer of 4Hb-TaS$_2$

As shown in Fig. S11, in the 1H layer of 4Hb-TaS$_2$, a ~15 mV gap coexists with the superconducting gap. This ~15 mV gap persists even when the superconducting gap is suppressed by the external magnetic field.

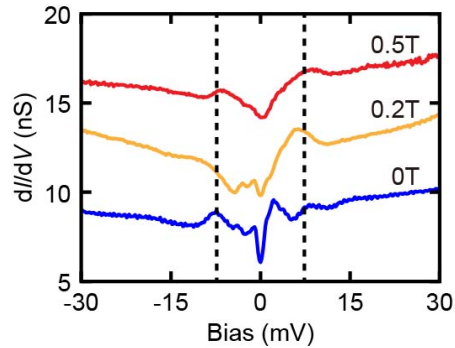

**Figure S11.** d*I*/d*V* spectra taken on the 1H layer of 4Hb-TaS$_2$ with different out-of-plane magnetic fields at 0.7 K. The vertical dashed lines indicate the gap-like feature. The spectra are vertically offset for clarity.